\documentclass[12pt]{article}
\usepackage{graphicx}
\usepackage{amssymb}
\usepackage{amsmath}

\begin{document}

\begin{center}
{\Large{\bf Amplification of Fluctuations in Unstable Systems with Disorder}} \\
\ \\
\ \\
by \\
Prabhat K. Jaiswal$^1$, Manish Vashishtha$^2$, Rajesh Khanna$^{2}$, and Sanjay Puri$^1$ \\
$^1$School of Physical Sciences, Jawaharlal Nehru University, New Delhi -- 110067, India. \\
$^2$Department of Chemical Engineering, Indian Institute of Technology Delhi,
New Delhi -- 110016, India.
\end{center}

\begin{abstract}
We study the early-stage kinetics of thermodynamically unstable systems with 
quenched disorder. We show analytically that the growth of initial fluctuations
is amplified by the presence of disorder. This is confirmed by numerical simulations
of {\it morphological phase separation} (MPS) in thin liquid films and {\it spinodal
decomposition} (SD) in binary mixtures. We also discuss the experimental implications
of our results.
\end{abstract}

\newpage

\section*{Introduction}

There has been intense research interest in the {\it kinetics of phase transitions} 
in systems which have been rendered thermodynamically unstable by a sudden change 
of parameters, e.g., temperature, pressure, magnetic field, etc. 
The subsequent far-from-equilibrium evolution of the system 
is usually nonlinear, and is characterized by complex \textit{spatio-temporal 
pattern formation}. The system approaches its new equilibrium state 
via the emergence and growth of domains enriched in the preferred phases. 
This nonequilibrium dynamics is often referred to as {\it phase-separation kinetics}
or \textit{domain growth} or \textit{coarsening} or \textit{phase ordering dynamics} 
\cite{pw09}. These processes occur over a wide range of length-scales and time-scales,
and are of great scientific and technological importance. 

In pure (disorder-free) systems, the coarsening domains are usually characterized 
by a divergent length scale which grows as a power-law, $L(t)\sim t^\phi$, 
where the exponent $\phi$ depends on the transport mechanism. For example, 
in the phase separation of an unstable binary mixture via diffusion, 
$L(t)\sim t^{1/3}$ in the late stages of evolution. This is known as the 
{\it Lifshitz-Slyozov} (LS) growth law. In recent works \cite{pre10,pccp10}, we 
have established that the LS growth law also describes the late-stage dynamics 
in the {\it morphological phase separation} (MPS) of an unstable thin liquid film
($< 100$ nm) on a solid substrate. During MPS, the film segregates into flat
regions (domains) and high-curvature regions (defects or hills). In these works,
we have also emphasized the analogies and differences between the kinetics of phase
separation in binary mixtures and thin films.

Of course, real experimental systems are never pure or disorder-free: they are 
invariably characterized by chemical and physical heterogeneity
\cite{kk2000,volo2007a,volo2007b,and91,and92,koch95,lip00,iei90,lla00,llop01}. Therefore, 
it is natural to investigate the effects of disorder on the domain growth process. 
For segregating $AB$ mixtures, coarsening is driven by interfaces between 
coexisting $A$-rich and $B$-rich domains. These interfaces are trapped by sites 
of quenched disorder, and subsequent growth occurs by thermally-activated 
hopping of interfaces over disorder traps \cite{hh85}. Thus, domain growth 
is drastically slowed down in the late stages, though the domain morphology 
is unaffected \cite{sp04,pcp91,ppr04,lmp10}. On the other hand, unstable liquid
films undergoing MPS segregate into regions with heights $h=h_m$ and $h=\infty$, i.e.,
there are no coexisting phases. The ``interface'' between the flat phase ($h=h_m$) 
and the high-curvature steepening phase ($h=\infty$) becomes sharper as time proceeds,
and is unaffected by the presence of disorder. Thus, the asymptotic regime of 
MPS with disorder is universal, and is characterized by the LS growth law \cite{pccp10}.

In this letter, we focus on the effect of disorder on the early stages of
phase separation, i.e., the growth of small fluctuations about a 
homogeneous initial state. As we discuss shortly, both binary mixtures
and thin films \cite{vsm93} are described by the {\it Cahn-Hilliard-Cook} (CHC) 
equation \cite{ch58,hc70} with appropriate free energies. The early-stage 
dynamics is described within the framework of a linear theory, usually 
referred to as CHC theory. In this letter, we report that growth in the 
CHC regime is strongly amplified by the presence of disorder. We present both 
analytical and numerical results to support this scenario. This should 
be contrasted with the late-stage dynamics, which is either slowed down 
(for binary mixtures) or unaffected (for thin films) 
by the presence of quenched disorder.

\section*{Dynamical Equations and Analytical Results}

The starting point of our discussion is the dimensionless form of the 
CHC equation, which describes the evolution of a conserved order parameter 
$\psi(\vec{r},t)$ at space-point $\vec{r}$ and time $t$ \cite{hh77}:
\begin{equation}
 \frac{\partial}{\partial t} \psi(\vec{r},t) = \vec{\nabla}\cdot 
 \left[M(\psi) \vec{\nabla}
 \left(\frac{\delta F[\psi]}{\delta\psi}\right)+\vec{\eta}(\vec{r},t)\right].
 \label{eq1:ch}
\end{equation}
In Eq.~(\ref{eq1:ch}), $M(\psi)$ is the mobility, and 
the Gaussian white noise $\vec{\eta}$ satisfies the fluctuation-dissipation relation:
\begin{eqnarray}
 \overline{\vec{\eta}(\vec{r},t)} &=& 0, \nonumber \\
 \overline{\eta_i(\vec{r}\,^\prime,t^\prime)\eta_j(\vec{r}\,^{\prime\prime},t^{\prime\prime})}
&=&2\epsilon M(\psi)\delta_{ij}\delta(\vec{r}\,^\prime-\vec{r}\,^{\prime\prime})\delta(t^\prime-t^{\prime\prime}).
 \label{eq2:dim-noise}
\end{eqnarray}
Here, the bars denote an averaging over the noise ensemble, and $\epsilon$ 
measures the strength of the noise. Further, $F[\psi]$ is the free-energy functional
which has the following form:
\begin{equation}
 F[\psi] = \int d\vec{r} \left[ f_a(\psi) + 
 \frac{1}{2}{\left(\vec{\nabla}\psi\right)}^2\right].
 \label{eq3:f}
\end{equation}
In Eq.~(\ref{eq3:f}), $f_a(\psi)$ denotes the local free energy which is 
parameterized by the disorder strength $a$. The second term on the RHS refers 
to the interfacial energy or surface tension. Replacing Eq.~(\ref{eq3:f}) 
in Eq.~(\ref{eq1:ch}), we obtain
\begin{equation}
 \frac{\partial}{\partial t} \psi(\vec{r},t) = \vec{\nabla}\cdot 
 \left[M(\psi) \vec{\nabla}
 \left\lbrace f_a^\prime (\psi)-\nabla^2\psi\right\rbrace+\vec{\eta}(\vec{r},t)\right].
 \label{eq4:ch2}
\end{equation}

The dimensionless Eqs.~(\ref{eq1:ch})-(\ref{eq4:ch2}) are obtained from the 
corresponding dimensional equations by using the natural scales of length, 
time, and order parameter \cite{pw09}. Before proceeding it is useful to 
discuss the functional form of $f_a(\psi)$ for the problems of interest here. For 
unstable thin films, $\psi$ is the height (usually denoted as $H$), and a typical 
form of $f_0(\psi)$ for the pure case ($a=0$) is 
\begin{equation}
 f_0(H)=-\frac{1}{6}\left[ \frac{1-R}{\left(H + D\right)^{2}}+ \frac{R}{H^{2}}\right].
 \label{eq:v-tf}
\end{equation}
This potential describes a thin film on a coated substrate. 
In Eq.~(\ref{eq:v-tf}), $R$ is the ratio of the effective Hamaker constants 
for the system, and $D$ is the dimensionless coating thickness. One can introduce 
the disorder in the coated potential through $R$ (chemically heterogeneous 
substrate) or $D$ (physically heterogeneous substrate). Here, we 
consider the case with chemical heterogeneity, so that $R$ is a $\vec{r}$-dependent 
random variable uniformly distributed in the interval $[R_{m}-a,R_{m}+a]$. 

For segregating binary mixtures, $\psi$ denotes the density difference of the 
two species: $\phi(\vec{r},t)=n_A(\vec{r},t)-n_B(\vec{r},t)$, where $n_\alpha$ is the
density of species $\alpha$. The local free energy for the disorder-free case is
modeled by a $\phi^4$-potential with a double-well structure:
\begin{equation}
 f_0(\phi) = -\frac{b\phi^2}{2} + \frac{\phi^4}{4},\quad b=1.
 \label{eq:v-bin}
\end{equation}
We model chemical heterogeneity by making $b$ a space-dependent random 
variable, which is uniformly distributed in $[1-a,1+a]$, i.e., the 
critical temperature varies from point to point. 

Let us now return to Eq.~(\ref{eq4:ch2}). To understand the dynamics of 
the early stages, we consider the growth of initial fluctuations, 
$\psi(\vec{r},t)=\psi_0 + \delta\psi(\vec{r},t)$. Here, $\psi_0$ is the 
homogeneous value of the order parameter in the unstable state. The 
linearization of Eq.~(\ref{eq4:ch2}) yields
\begin{equation}
 \frac{\partial}{\partial t} \delta\psi(\vec{r},t) = M(\psi_0)
 \nabla^2 (-\alpha - \nabla^2)\delta\psi+\vec{\nabla}\cdot\vec{\eta}(\vec{r},t),
 \label{eq5:tr}
\end{equation}
where we have introduced the parameter $\alpha\equiv -f_a^{\prime\prime}(\psi_0)$.
We neglect terms involving the spatial derivatives of the disorder variable, as
these are only relevant on microscopic scales. (We will subsequently average all
physical quantities over the disorder distribution.) The Fourier transformation of
Eq.~(\ref{eq5:tr}) (with wave-vector $\vec{k}$) gives
\begin{equation}
 \frac{\partial}{\partial t} \delta\psi(\vec{k},t) = M(\psi_0)
  k^2(\alpha - k^2 )\delta\psi(\vec{k},t)-i\,\vec{k}\cdot\vec{\eta}(\vec{k},t),
 \label{eq6:tk}
\end{equation}
which has the following solution:
\begin{eqnarray}
 \delta\psi(\vec{k},t) &=& \exp\left[M(\psi_0)k^2 
 (\alpha - k^2 ) t\right] \delta\psi(\vec{k},0) \nonumber \\ 
 && - i\,\vec{k}\cdot  \int_0^{t} dt^\prime \exp
\left[M(\psi_0)k^2(\alpha-k^2)(t-t^\prime)\right]
 \vec{\eta}(\vec{k},t^\prime) .
 \label{eq7:tksol}
\end{eqnarray}
To obtain Eq.~(\ref{eq7:tksol}), we have set $M(\psi)\simeq M(\psi_0)$ in 
Eq.~(\ref{eq2:dim-noise}), which is valid at early times.
Clearly, averaging over the noise ensemble yields
\begin{equation}
 \overline{\delta\psi}(\vec{k},t) = \exp\left[M(\psi_0)k^2 
 (\alpha - k^2 ) t\right] \delta\psi(\vec{k},0) .
 \label{eq8:tn}
\end{equation}

The parameter $\alpha$ is spatially uniform for a chemically 
homogeneous system. For the disordered system, we have the local free 
energy
\begin{equation}
 f_a(\psi) = f_0(\psi) + xf_1(\psi),
 \label{eq:V}
\end{equation}
where $x$ is a uniformly-distributed random variable in the interval 
$[-a,a]$. For the disorder types we consider here, Eq.~(\ref{eq:V}) is 
exact. Otherwise, it represents the first two terms of a Taylor expansion 
in the disorder $x$. Thus, $\alpha (x)= -f_0^{\prime\prime}(\psi_0) 
-xf_1^{\prime\prime}(\psi_0)$.

Finally, integrating Eq.~(\ref{eq8:tn}) over the uniform disorder distribution 
with the probability $P(x)=1/(2a)$ gives
\begin{equation}
 \widetilde{\delta\psi}(\vec{k},t)=\int_{-a}^a dx\, \frac{1}{2a} \exp\left[
 M(\psi_0)k^2  \left\{\alpha(x) - k^2 \right\} t\right] \delta\psi(\vec{k},0).
 \label{eq:tint}
\end{equation}
Some simple algebra results in the following expression for the disorder-averaged 
growth of fluctuations:
\begin{equation}
 \widetilde{\delta\psi}(\vec{k},t) = \exp\left[M(\psi_0)k^2\left\lbrace-f_0^{\prime\prime}(\psi_0)
-k^2\right\rbrace t\right] \frac{\sinh[M(\psi_0)f_1^{\prime\prime}(\psi_0)k^2a\,t]}
 {M(\psi_0)f_1^{\prime\prime}(\psi_0)k^2a\,t} \,\delta\psi(\vec{k},0).
 \label{eq:theta}
\end{equation}
The factor $\sinh(ct)/(ct)$ determines the effect of disorder on the 
order-parameter evolution. This factor tends to 1 as $a \to 0$ or 
$k \to 0$ or $t \to 0$. However, the growth of initial fluctuations can be 
strongly amplified for nonzero $a$ as $\sinh(ct)$ grows exponentially with $t$.
 
Finally, we evaluate the {\it time-dependent structure factor}, which is probed in 
scattering experiments using, e.g., light, neutrons. This is defined as follows:
\begin{equation}
 S(\vec{k},t) = \langle\delta\psi(\vec{k},t)\delta\psi^*(\vec{k},t)\rangle ,
 \label{eq9:sk}
\end{equation}
where the angular brackets indicate an averaging over thermal noise and 
independent initial conditions. Using the expression for $\delta\psi(\vec{k},t)$ 
from Eq.~(\ref{eq7:tksol}), we obtain 
\begin{eqnarray}
 S(\vec{k},t) = &A&\!\!\!\!V\exp\left[2M(\psi_0)k^2  \left(\alpha - k^2 \right) t\right] 
\nonumber \\
&+&\!\!\!\! \frac{\epsilon V}{\left(\alpha - k^2 \right)}
\left\{\exp\left[2M(\psi_0)k^2  \left(\alpha - k^2 \right) t\right]-1\right\} .
\label{eq10:sk2}
\end{eqnarray}
To obtain Eq.~(\ref{eq10:sk2}), we assume that $\langle\delta\psi(\vec{r},0)
\delta\psi^*(\vec{r^\prime},0)\rangle = A \delta(\vec{r}-\vec{r^\prime})$, 
where $A$ is the amplitude of initial fluctuations in the order parameter. 
Further, $V$ denotes the volume of the system.
The corresponding disorder-averaged structure factor is 
\begin{eqnarray}
 \widetilde{S}(\vec{k},t) = &A&\!\!\!\!V\exp\left[2M(\psi_0)k^2  \left\{-f_0^{\prime\prime}(\psi_0) 
- k^2 \right\} t\right]\frac{\sinh[2M(\psi_0)f_1^{\prime\prime}(\psi_0)k^2a\,t]}
 {2M(\psi_0)f_1^{\prime\prime}(\psi_0)k^2a\,t} 
\nonumber \\
&-&\!\!\!\! \frac{\epsilon V}{2af_1^{\prime\prime}(\psi_0)}\int_{-f_0^{\prime\prime}(\psi_0)
+af_1^{\prime\prime}(\psi_0)-k^2}^{-f_0^{\prime\prime}(\psi_0)-af_1^{\prime\prime}(\psi_0)-k^2} dy
\left\{\frac{\exp\left[2M(\psi_0)k^2  y t\right]-1}{y}\right\} .
\label{eq11:skd}
\end{eqnarray}

\section*{Numerical Results}

Next, we describe our numerical results for the chosen systems. 
For MPS in a thin film, the appropriate model is 
Eq.~(\ref{eq4:ch2}) with $\psi\equiv H$ and the local free energy:
\begin{equation}
 f_a(H)=f_0(H)+xf_1(H), \quad f_1(H) = \frac{1}{6}\left[\frac{1}{(H+D)^2}
-\frac{1}{H^2}\right] ,
\label{eq:1n}
\end{equation}
where $f_0(H)$ is defined in Eq.~(\ref{eq:v-tf}). Here, we consider the case 
with mobility $M(H)=H^3$ (corresponding to Stokes flow with no slip) and
$\epsilon=0$, i.e., without noise. 
We numerically solve Eq.~(\ref{eq4:ch2}) in $d=1$ starting with a small-amplitude 
($\simeq 0.01$) random perturbation about the mean film thickness, $H_0 = 1$. 
The system size is $n\bar{L}_{M}$, where $\bar{L}_M$ is the 
dominant wavelength for $R=R_m$ ($n$ ranges from 16 to several thousands). 
We apply periodic boundary conditions at the end points. 
A 512-point grid per $\bar{L}_{M}$ was found to be 
sufficient when central-differencing in space with half-node interpolation 
was combined with {\it Gear's algorithm} for time-marching, which is 
convenient for stiff equations. The parameters $D=0.5$, $R_{m}=-0.1$ and $a$ 
were chosen so that the film is spinodally unstable at $H = 1$ 
(i.e., $\alpha > 0$) for all values of $R$. 

For segregation in a disordered binary mixture, we consider Eq.~(\ref{eq4:ch2}) with 
$\psi\equiv \phi$ and the free energy: 
\begin{equation}
 f_a(\phi) = f_0(\phi) + x f_1(\phi), \quad f_1(\phi) = -\frac{\phi^2}{2} ,
\label{eq:2n}
\end{equation}
where $f_0(\phi)$ is defined in Eq.~(\ref{eq:v-bin}). We consider the case 
with $M(\phi)=1$ and $\epsilon=0$. We numerically solve Eq.~(\ref{eq4:ch2}) via an 
Euler-discretization scheme on a $d=2$ lattice of size $512^2$. The mesh sizes 
in space and time are $\varDelta x=0.5$ and $\varDelta t=0.001$, respectively. The
spatial mesh size is small enough to resolve the interface region between $A$-rich
($\phi = +1$) and $B$-rich ($\phi = -1$) domains. The interface thickness $\xi \simeq
\sqrt{2}$ in our dimensionless units. The initial condition for a run consists of
small-amplitude ($\simeq 0.01$) fluctuations about $\phi_0=0$, corresponding to a
critical $AB$ mixture with $50\% A$ and $50\% B$. 

\begin{figure}[!htbp]
\centering
    \begin{tabular}{cc}
      \includegraphics*[width=0.80\textwidth]{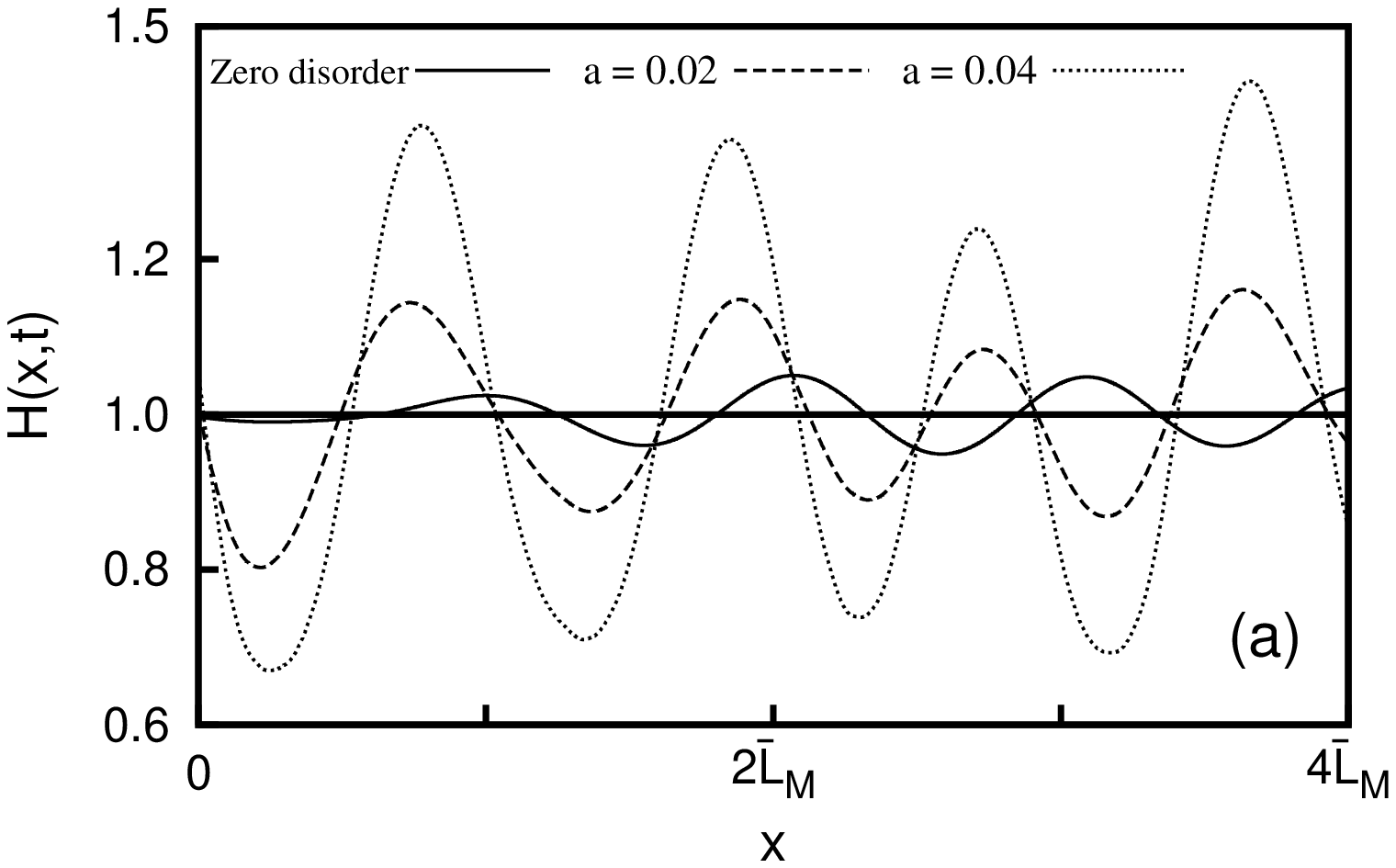} \\ \\
      \includegraphics*[width=0.60\textwidth]{figure1b.eps}
    \end{tabular}
\caption{Early stages of morphological phase separation (MPS) in an unstable thin film
with quenched disorder.
(a) Height profiles [$H(x,t)$ vs. $x$] at $t=4400$ for different disorder amplitudes. The
simulation details are provided in the text. (b) Plot of $\langle \delta H(x,t)^2
\rangle$ vs. $t$ for the height profiles shown in (a).}
\label{fig:fig1}
\end{figure}

First, we present results for MPS in
films with chemical heterogeneity. Recall that we are interested in the effect 
of disorder on the early stages of growth, i.e., the CHC regime. The height 
profiles at a fixed time ($t=4400$) for various disorder strengths are shown 
in Fig.~\ref{fig:fig1}(a). The initially random perturbations grow exponentially 
in time, with fastest growth for the case with maximum disorder ($a=0.04$). The 
growing fluctuations rapidly select the most unstable wavelength, $\lambda=
\bar{L}_M$ -- the location of the zero crossings is unaffected by the disorder. 
However, the peaks of the profiles are strongly amplified due to the factor 
$\sinh(ct)/(ct)$ in Eq.~(\ref{eq:theta}). This amplification may be quantitatively
characterized by studying the statistical properties of the height profile. In 
Fig.~\ref{fig:fig1}(b), we plot $\widetilde{C} (0,t)=\langle\delta H(x,t)^2\rangle$ vs. 
$t$. This quantity is related to the structure factor in Eq.~(\ref{eq9:sk}) as
\begin{equation}
 \widetilde{C}(0,t) = \int \frac{d\vec{k}}{(2\pi)^d}\, \widetilde{S}(\vec{k},t) ,
\label{eq:n3}
\end{equation}
where $d$ is the dimensionality. As expected, $\widetilde{C}(0,t)$ grows fastest
for the case with maximum disorder. 

\begin{figure}[!htbp]
\centering
\includegraphics*[width=0.95\textwidth]{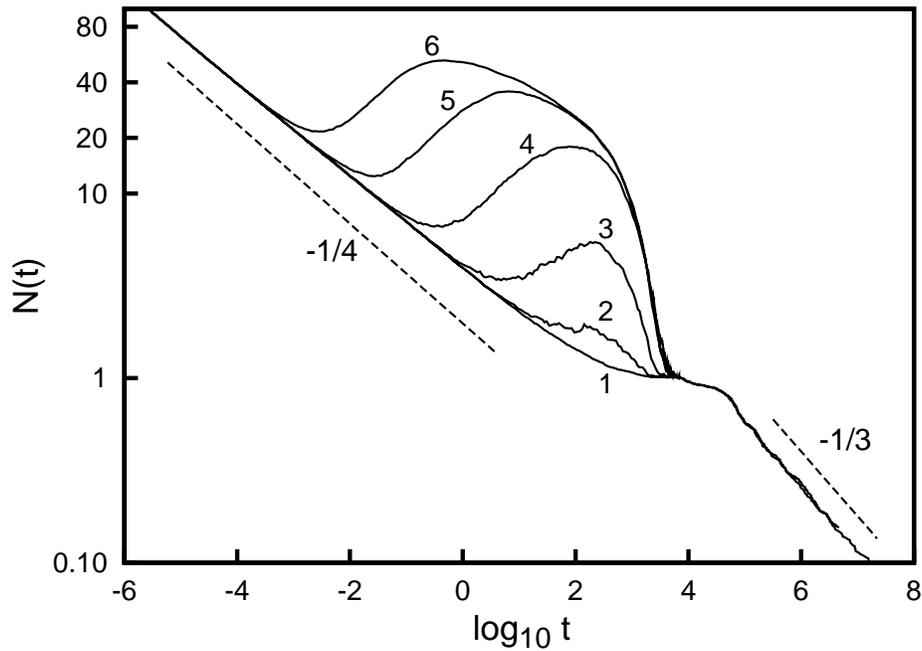}
\caption{Variation of the number density of defects with time [$N(t)$ vs. $t$] for
MPS in an unstable thin film. Curves $1$ to $6$ present results for disorder amplitudes
$a=0,0.0001, 0.0004,
0.002, 0.01$ and $0.04$, respectively. The dashed lines with slopes of $-1/4$ and $-1/3$
denote exponents for the early and late stages, respectively \cite{pre10,pccp10}.} 
\label{fig:fig2}
\end{figure}

The initial exponential growth of the profiles in the CHC regime is followed by a
nonlinear saturation which results in the formation of domains of the flat phase,
separated by defects of the high-curvature phase. These domains grow via 
the diffusive transport of material. The CHC regime and the domain growth regime 
are both shown in Fig.~\ref{fig:fig2}, where we plot the number of 
defects (hills) $N(t)$ vs. $t$. The characteristic length scale is $L(t) 
\sim N(t)^{-1}$. The CHC regime is followed by an {\it intermediate stage} 
where disorder sites trap the emergent interfaces. In this regime, the more 
disordered systems show slower growth. The intermediate stage is very sensitive
to the presence of even small amounts of disorder. In the {\it late stage}, the domain 
growth is unaffected by disorder. In this regime, the interfaces (between 
$H=H_m$ and $H=\infty$) have steepened too much to be captured by the local 
disorder \cite{pccp10}. Thus, this regime shows universal LS growth [$N(t)
\sim t^{-1/3}$ or $L(t)\sim t^{1/3}$]. This result has important experimental 
implications, viz., it is not necessary to work with specially-prepared
extra-pure substrates, at least in the context of domain growth studies.

\begin{figure}[!htbp]
\centering
\includegraphics*[width=0.95\textwidth]{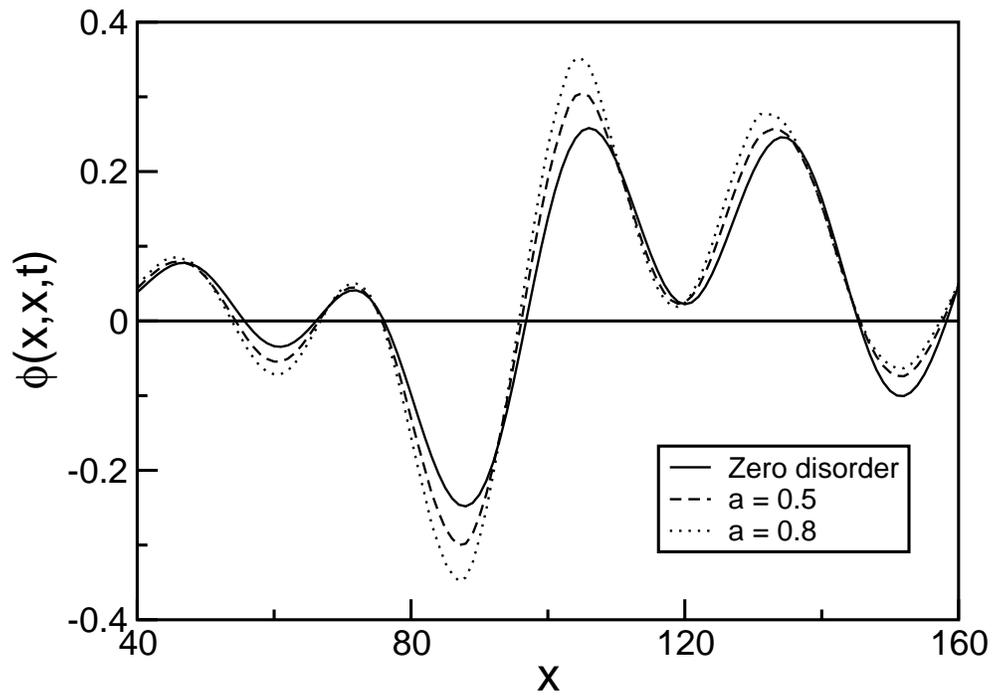}
\caption{Early stages of spinodal decomposition (SD) in an unstable disordered binary mixture in
$d=2$. The plot shows the variation of the order parameter along the diagonal [$\phi (x,x,t)$
vs. $x$] at $t=20$ for different disorder amplitudes. The simulation details are given in the
text.}
\label{fig:fig3}
\end{figure}

Next, we present our results for the early stages of spinodal decomposition (SD)
in $d=2$ binary mixtures. In Fig.~\ref{fig:fig3}, 
we show the variation of the order parameter along a diagonal of the lattice 
($y=x$). We plot profiles  for the pure ($a=0$) and disordered ($a=0.5,0.8$) 
cases at $t=20$. Again, we see that there is an amplification of fluctuations for 
higher disorder values, though the selected wavelength does not change.

\begin{figure}[!htbp]
\centering
\includegraphics*[width=0.95\textwidth]{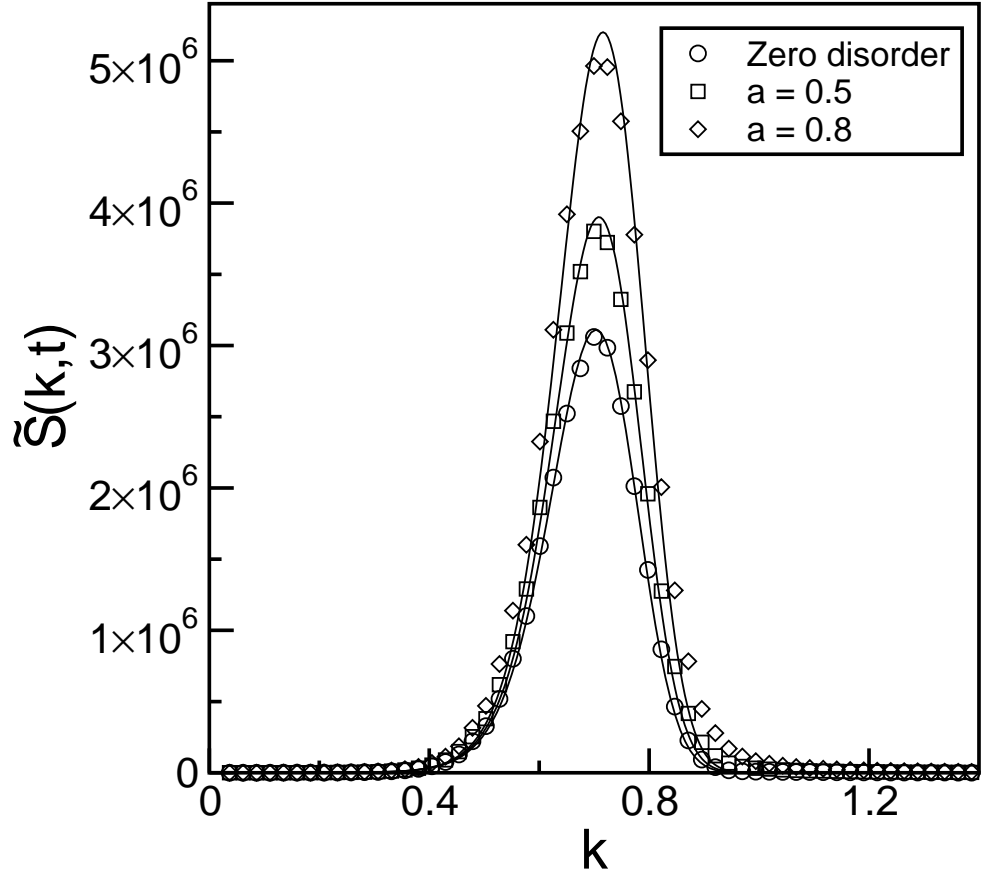}
\caption{Plot of the spherically-averaged structure factor [$\widetilde{S}(k,t)$ vs. $k$]
at $t=20$ for different disorder amplitudes. The structure factor is obtained as an average
over 50 independent runs. The initial condition for each run consists of a different order
parameter and disorder configuration. The solid lines denote the expression in
Eq.~(\ref{eq11:skd}) with $\epsilon = 0$.}
\label{fig:fig4}
\end{figure}

Finally, we study the structure factor $S(k,t)$ which provides a quantitative measure 
of the phase-separating morphology. In Fig.~\ref{fig:fig4}, we plot $S(k,t)$ vs. $k$ 
at $t=20$ for the pure and disordered cases shown in Fig.~\ref{fig:fig3}. In the 
early stages or the CHC regime, the structure factor grows
exponentially in time but the peak position 
does not shift, i.e.,  there is no change in the characteristic scale. We see that 
the fastest growth occurs for the highest disorder amplitude. The solid lines in 
Fig.~\ref{fig:fig4} denote the expression in Eq.~(\ref{eq11:skd}) with 
$\epsilon=0$. Our numerical results are seen to be in excellent agreement with theory. 

In the late stages of phase separation, disorder has the opposite effect, drastically 
slowing down domain growth \cite{sp04,pcp91,ppr04,lmp10}. As stated earlier, this is
because interfaces between coexisting $A$-rich and $B$-rich domains are trapped by
sites of quenched disorder. This should be contrasted with the universal late-stage
scenario for segregating films, as described above.

\section*{Summary and Discussion}

In summary, we have studied the effect of quenched disorder on the early stages of 
evolution in two important systems: unstable thin films and 
phase-separating mixtures. Both these systems are described by the Cahn-Hilliard-Cook 
(CHC) equation with appropriate choices of the free energy. We find that disorder 
amplifies the exponential growth of fluctuations in the early CHC regime in both 
systems, and we have obtained a simple analytical expression for the amplification 
factor. In the asymptotic regime, there is a major difference between disordered domain 
growth in these two systems, viz., films undergoing MPS are unaffected by disorder,
whereas segregating mixtures are drastically slowed down by disorder. Our results have 
important experimental consequences, and we hope that they will be subjected to 
experimental tests.

\newpage

\newpage

\newpage

\newpage


\newpage
\begin{thebibliography}{99}

\bibitem{pw09} S. Puri and V. K. Wadhawan (eds.), {\it Kinetics of Phase
Transitions}, CRC Press, Boca Raton, Florida (2009).

\bibitem{pre10} R. Khanna, N. K. Agnihotri, M. Vashishtha, A. Sharma, 
P. K. Jaiswal, and S. Puri, Phys. Rev. E {\bf 82}, 011601 (2010).

\bibitem{pccp10} M. Vashishtha, P. K. Jaiswal, R. Khanna, S. Puri, and 
A. Sharma, Phys. Chem. Chem. Phys. {\bf 12}, 12964 (2010).

\bibitem{kk2000} K. Kargupta, R. Konnur, and A. Sharma, 
Langmuir {\bf 16}, 10243 (2000).

\bibitem{volo2007a} P. Volodin and A. Kondyurin, Journal of Physics D: 
Applied Physics {\bf 41}, 065306 (2008).

\bibitem{volo2007b} P. Volodin and A. Kondyurin, Journal of Physics D: 
Applied Physics {\bf 41}, 065307 (2008).

\bibitem{and91} M. O. Robbins, D. Andelman, and J.-F. Joanny, 
Phys. Rev. A {\bf 43}, 4344 (1991).

\bibitem{and92} J. L. Harden and D. Andelman, Langmuir {\bf 8}, 2547 (1992).

\bibitem{koch95} W. Koch, S. Dietrich, and S. Napiorkowski, 
Phys. Rev. E {\bf 51}, 3300 (1995).

\bibitem{lip00} R. Lipowsky, P. Lenz, and P. S. Swain, Colloids and 
Surfaces A: Physicochemical and Engineering Aspects {\bf 161}, 3 (2000). 

\bibitem{iei90} H. Ikeda, Y. Endoh and S. Itoh, Phys. Rev. Lett.
{\bf 64}, 1266 (1990).

\bibitem{lla00} V. Likodimos, M. Labardi and M. Allegrini, Phys. Rev. B {\bf 61},
14440 (2000).

\bibitem{llop01} V. Likodimos, M. Labardi, X. K. Orlik, L. Pardi, M. Allegrini,
S. Emonin and O. Marti, Phys. Rev. B {\bf 63}, 064104 (2001).

\bibitem{hh85} D. A. Huse and C. L. Henley, Phys. Rev. Lett. {\bf 54}, 2708 (1985). 

\bibitem{sp04} S. Puri, Phase Transitions {\bf 77}, 469 (2004).

\bibitem{pcp91} S. Puri, D. Chowdhury, and N. Parekh, J. Phys. A {\bf 24}, L1087 (1991);
S. Puri and N. Parekh, J. Phys. A {\bf 25}, 4127 (1992).

\bibitem{ppr04} R. Paul, S. Puri, and H. Rieger, Europhys. Lett. {\bf 68}, 881 (2004);
Phys. Rev. E {\bf 71}, 061109 (2005).

\bibitem{lmp10} E. Lippiello, A. Mukherjee, S. Puri, and M. Zannetti, 
Europhys. Lett. {\bf 90}, 46006 (2010).

\bibitem{vsm93} V. S. Mitlin, J. Colloid Interface Sci. {\bf 156}, 491 (1993).

\bibitem{ch58} J. W. Cahn and J. E. Hilliard, J. Chem. Phys. {\bf 28}, 
258 (1958).

\bibitem{hc70} H. E. Cook, Acta Metall. {\bf 18}, 297 (1970).

\bibitem{hh77} P.C. Hohenberg and B.I. Halperin, Rev. Mod. Phys.
\textbf{49}, 435 (1977).

\end{thebibliography}
\end{document}